\documentclass[%
 aip,
 amsmath,amssymb,
 reprint,%
]{revtex4-1}

\usepackage{graphicx}% Include figure files
\usepackage{dcolumn}% Align table columns on decimal point
\usepackage{bm}% bold math
\usepackage{multirow}
\usepackage[utf8]{inputenc}
\usepackage{array}
\usepackage[T1]{fontenc}
\usepackage{mathptmx}
\usepackage{subfigure}
\usepackage{upgreek}
\usepackage{xcolor}
\usepackage{soul}
\usepackage[colorlinks=true,bookmarks=false,citecolor=blue,urlcolor=blue]{hyperref} %pdflatex

\begin{document}

\preprint{AIP/123-\textbf{Q}ED}

\title{Towards Scalable, Energy-Efficient and Ultra-Fast Optical SRAM}
% Force line breaks with \\
\author{Ramesh Kudalippalliyalil}
\altaffiliation[]{Contributed equally to this work}
\author{Sujith Chandran}%
\altaffiliation[]{Contributed equally to this work}
\author{Ajey P. Jacob}
 \altaffiliation[Authors to whom correspondence should be addressed: ]{ajey@isi.edu and akjaiswal@isi.edu (Contributed equally to this work).}
  %\altaffiliation[]{Equal Contributors.}
\author{Akhilesh Jaiswal}
 \altaffiliation[Authors to whom correspondence should be addressed: ]{ajey@isi.edu and akjaiswal@isi.edu (Contributed equally to this work).}
  % \altaffiliation[]{Equal Contributors}
 \affiliation{$^1$Information Sciences Institute (ISI), University of Southern California (USC), Marina Del Rey, CA 90292} %https://sites.usc.edu/asiclab/}

\date{\today}% It is always \today, today,
             %  but any date may be explicitly specified

\begin{abstract}
Optical static random access memory (O-SRAM) is one of the key components required for achieving the goal of ultra-fast, general-purpose optical computing. We propose and design a novel O-SRAM using fabrication-friendly photonics device components such as cross-coupled micro-ring resonators and photodiodes. Based on the chosen photonic components, the memory operates at a speed of 20 Gb/s and requires ultra-low static (switching) energy of $\sim16.7~$aJ/bit ($\sim1.04$~pJ/bit) to store a single bit. The footprint of the bit cell is $\sim2400~\mu$m$^2$. The proposed O-SRAM can be configured in a 2D memory array by replicating the bit-cells along rows and columns for creating ultra-large scale on-chip optical memory sub-system. Such manufacturing-friendly, large-scale optical-SRAM could form the underlying memory backbone for photonics integrated circuits  with wide applications in novel computing and networking.
  
\end{abstract}

\maketitle

%\begin{quotation}
 
% \end{quotation}

%%%%%%%%%%%%%%%%%%%%%%%%%%%%%%%%%%%%%
%\section{Introduction}

The remarkable scalability of the electronic transistor has driven decades of improvements in power, performance and footprint metrics of digital computing systems \cite{khan2018science}. However, the state-of-the-art electronic platforms fail to provide the required computing speed, throughput, and energy-efficiency, demanded by emerging class of data intensive applications such as artificial intelligence, machine learning \cite{ahmed2020democratization}, extreme-scale simulations \cite{donofrio2009energy}. \textit{etc}. 
As a result, use of optical state variable for memory, computing and interconnects \cite{ambs2010optical, sawchuk1984digital}, has emerged as a  promising alternative to the electronic charge-based computing platforms that are limited by slow and energy-expensive data access and lower electrical bus bandwidth \cite{mckee2004reflections}.
%employing optical memories and optically interconnected memory-processor architectures, has been emerged as a potential solution to surpass the fundamental memory-wall bottleneck \cite{mckee2004reflections} of existing charge-based computing platforms, due to the long access time, energy dissipation and limited electrical bus bandwidth. 
There have been several demonstrations of large bandwidth memory-processor optical links \cite{maniotis2013optical,kodi2005design,brunina2012energy,yin2010experimental} reported in the past few years. However, high-speed, energy-efficient and compact photonic storage with true static random-access characteristics is still demanding to form the ultimate underlying memory sub-system for general-purpose optical computing. Various methods have been investigated for all-optical static random access memory (SRAM) cells, including semiconductor optical amplifier-Mach-Zehnder interferometer (SOA-MZI) configurations \cite{liu2006packaged,pleros2008optical,pitris2016wdm,tsakyridis201910}, SOA-based ring lasers \cite{li2009optical}, III-V photonic crystal cavities \cite{alexoudi2016iii} and optomechanical cavities \cite{dong2015nano}. SOA-MZI based optical SRAM bit cell reported in \cite{tsakyridis201910} exhibit a speed of 10 Gbps, however, these devices are not compatible for large scale chip-level memory integration as it requires a large footprint in the order of mm$^2$. On the other hand, photonic crystal cavities \cite{alexoudi2016iii} and nano-optomechanical cavities \cite{dong2015nano} can be integrated in a smaller footprint ($\sim100~\mu$m$^2$), but require major modifications to the existing foundry processes involving material or device integration aspects.

In this letter, we present a novel optical memory element based on cross-coupled optical devices - photodiodes (PDs) and microring resonator (MRR) modulators/switches.  Furthermore, the data can be selectively written or read to/from the optical memory element through another set of PDs and MRRs enabling random access functionality. This, in turn, leads to the construction of optical SRAM with striking functional similarities to electrical SRAMs, including complementary data storage and differential read-out of data \cite{mathur2020buried}. Furthermore, the proposed optical-SRAM (O-SRAM) can be arranged in an array-like fashion for creating large-scale on-chip optical storage. Advantageously, the use of well-known optical devices makes our proposal amenable to large-scale manufacturing in the existing silicon photonics foundry process. %Our simulation results suggest that the proposed O-SRAM can operate at a speed of $\sim10$~Gbps with ultra-low static (aJ/bit) and switching ($\sim1$ pJ/bit) energy consumption (excluding the thermal tuning). Furthermore, the overall footprint of the device can be as small as $\sim200~\mu$m$^2$. %The proposed O-SRAM bit-cell can be used to create an ultra-fast, ultra-large-scale on-chip memory system as needed for achieving the holy grail of general-purpose computing.

\begin{figure*}[t]
\centering
\includegraphics[scale=.35]{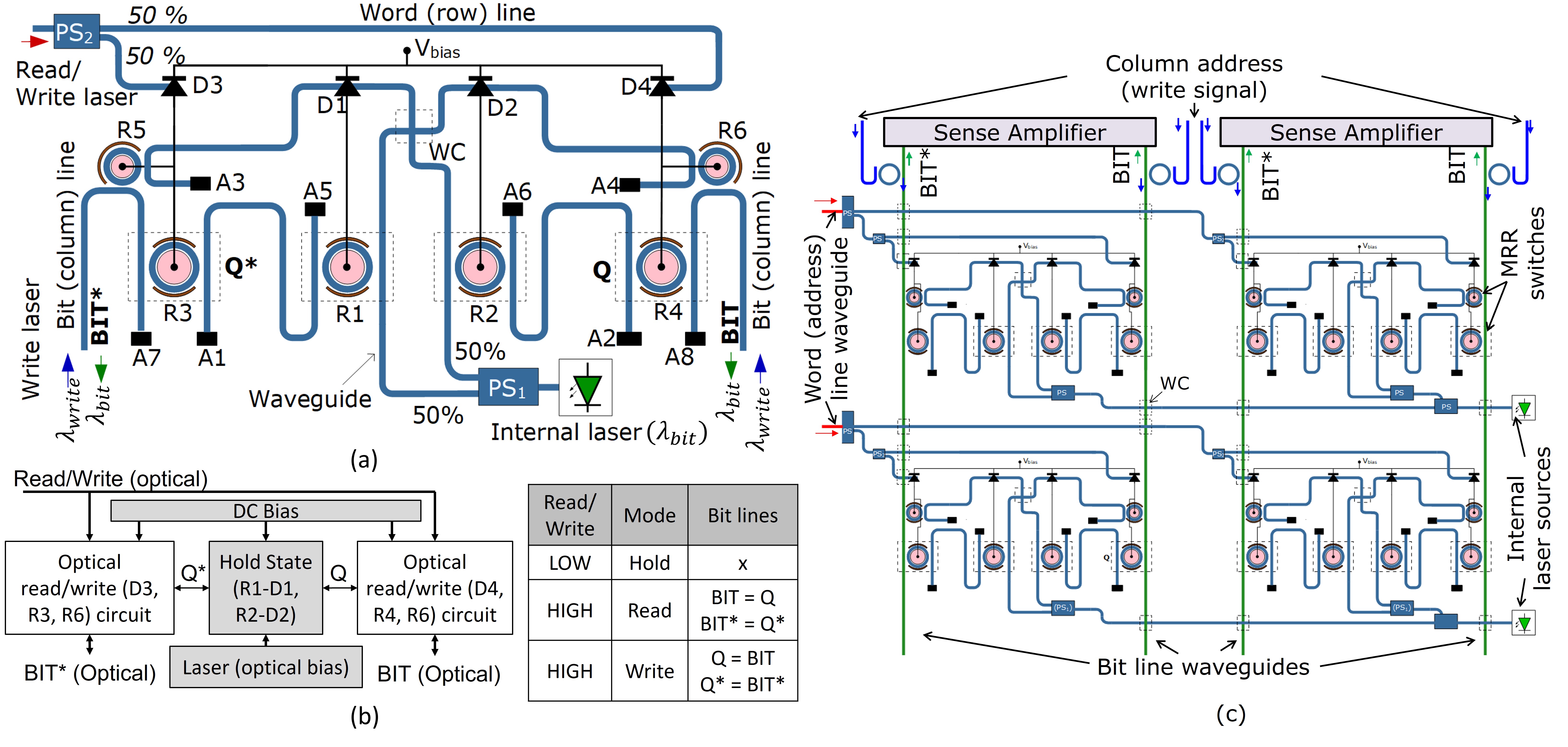}
\caption{(a) The proposed O-SRAM cell layout, along with (b) its block diagram and truth table. R1-R6: active MRR switches, D1-D4: photodiodes, PS1-PS2: power splitters, A1-A8: passive light absorber, WC: waveguide crossing. The electrical wires and optical waveguides  are represented by thin black lines and thick blue lines respectively. (c) O-SRAM cells arranged in a 2D memory array. }
\label{layout1}
\end{figure*}

Fig. \ref{layout1}(a) shows the schematic layout of our proposed O-SRAM cell, where R1-R6 are MRR based active optical switches, D1-D4 are photodiodes (PDs) coupled to integrated optical waveguides, PS1-PS2 are optical power splitters, and A1-A8 are passive optical absorbers to reduce reflections from unwanted output ports.  A block diagram representation of the O-SRAM is shown in Fig. \ref{layout1}(b) along with its truth table. The circuit is optically biased by an internal laser source (wavelength $\lambda_\text{bit}$) and electrically biased by a DC voltage source ($V_\text{bias}$).
The MRR switches R1 and R2, and feedback photodiodes D1 and D2 form the internal optical latch. The optical outputs \textbf{Q*} and \textbf{Q} represents the present state of the memory. While, the other MRRs (R3-R6) and PDs (D3 and D4) constitute the read/write access circuit.
The MRR switches are designed on a high speed p-n (carrier depletion) phase shifters. A detailed working of the circuit is discussed below.
The O-SRAM bit-cells can be arranged in a 2D array fashion, as shown in Fig. \ref{layout1}(c), by replicating the bit-cells in rows and columns for creating large-scale optical memory sub-system. The rows share `wordline' waveguides, while columns share BIT and BIT* waveguides.

%%%%%

%%%%%
%\begin{figure}[t]
%\centering
%\begin{subfigure}[]
%{\includegraphics[scale=.255]{Figures/layout.PNG}\label{layout}}
%\end{subfigure}
%\begin{subfigure}[]
%{\includegraphics[scale=.34]{Figures/blockdiagram.png}\label{block}}
%\end{subfigure}
%\caption{The proposed O-SRAM cell layout (a), along with a block diagram and truth table (b). R1-R6: active MRR switches, D1-D4: photodiodes, PS1-PS2: power splitters, A1-A8: light absorbing elements, WC: waveguide crossing, Q(Q*): hold state of the memory at $\lambda_\text{bit}$, BIT(BIT*) represent the read output (at $\lambda_\text{bit}$) or write input (at $\lambda_\text{write}$) of the memory. The electrical wires and optical waveguides  are represented by thin black lines and thick blue lines respectively.}
%\label{osram1}
%\end{figure}

As shown in Fig. 1(a), the internal bias laser ($\lambda_\text{bit}$) is coupled to a 50:50 power splitter (PS1) that feeds two identical MRR switches, R1 and R2, which in turn are controlled by the corresponding feedback photodiodes D1 and D2, respectively. The through-port of R1 (R2) drives D2 (D1). The drop ports of R1 and R2 represent the output states \textbf{Q*} and \textbf{Q} respectively. The ON state of D1 implies that R1 is resonant to the input light at $\lambda_\text{bit}$ and transfers maximum optical power to its drop-port, resulting in a logic HIGH at \textbf{Q*}. This also implies that the light output at the through-port of R1 is insufficient to turn ON D2, causing R2 to be not in resonance with the incoming light. As a result, R2 delivers power to D1 and keeps the circuit in a stable state (\textbf{Q*} = HIGH and \textbf{Q} = LOW) by keeping R1 ON. This state continues as long as the electrical and optical bias signals are applied to the memory element allowing a static storage of optical data. Thus, the cross-coupled MRR-PD system (R1, D2 \& R2, D1)  forms an optical latch or optical storage unit constituting an optical SRAM bit cell. 
The read/write part of the O-SRAM cell (R3 and R5 controlled by D3 \& R4 and R6 controlled by D4) remain OFF during the hold state unless the O-SRAM cell is accessed (or selected) for read/write operation by passing light through the `wordline' waveguide. Note, the O-SRAM would be initialized by a write operation (described below) to ensure either Q = HIGH, \textbf{Q*} = LOW (digital 1) or \textbf{Q} = LOW, \textbf{Q*} = HIGH (digital 0). Thus, the condition that \textbf{Q} and \textbf{Q* }can simultaneously be HIGH or LOW is obviated (similar to electrical-SRAMs that have a metastable state but are never encountered in normal memory operations).

In order to read the state of the memory, a read/write laser is first enabled in the `wordline' waveguide. This first drives D3 and D4 to the ON state; subsequently R3 and R4 are driven in resonant to $\lambda_\text{bit}$ (MRRs R1-R4 are identical). Thus, the \textbf{Q} (input of R4) and \textbf{Q*} (input of R3) will transfer to the drop ports of R3 and R4, respectively. This differential (complementary) optical outputs are then transferred to BIT and BIT* waveguides (R5 and R6 are designed to be non-resonant to $\lambda_\text{bit}$ irrespective of the switching voltage), which can then be fed to the peripheral sensing circuits for reading the stored data in the O-SRAM bit-cell. 

To understand the memory write operation, assume the memory is initially in \textbf{Q} = LOW and \textbf{Q*} = HIGH state. This means that D2 is in OFF state and R2 is off-resonant to $\lambda_\text{bit}$, while D1 is in ON state and R1 is in resonant to $\lambda_\text{bit}$. In order to flip the memory output state, D2 needs to be triggered with an external optical signal. The write operation is initiated by activating D3-D4 photodiodes by enabling the READ/WRITE laser in the `wordline'. An external optical write pulse (at $\lambda_\text{write}$) is then applied at the BIT waveguide, reaching D2 through R6 (resonant to $\lambda_\text{write}$). This process turns ON D2 and subsequently drives R2 in resonance to $\lambda_\text{bit}$. The resonant R2, turns away the light feeding D1 from its through-port to the drop port. In other words, D1 turns OFF and makes R1 non-resonant to $\lambda_\text{bit}$. This keeps D2 in ON state, leading to \textbf{Q} = HIGH and \textbf{Q*} = LOW, i.e., the state of the memory has switched as dictated by the write operation.

%\section{Simulations and Discussion}
\begin{figure*}[t]
\centering
\begin{subfigure}[]
{\includegraphics[scale=0.41]{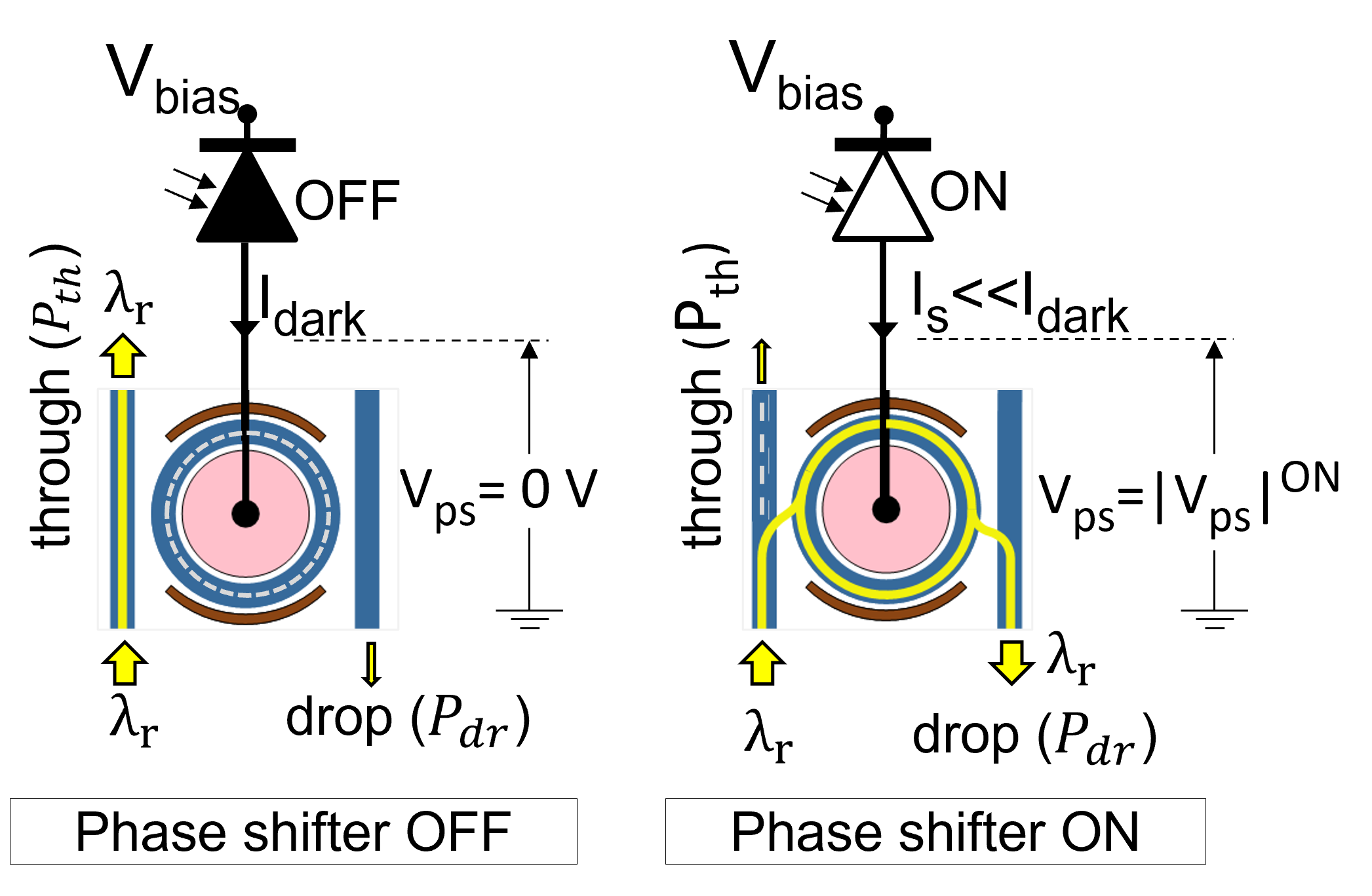}\label{mrr}}
\end{subfigure}
\begin{subfigure}[]
{\includegraphics[scale=0.32]{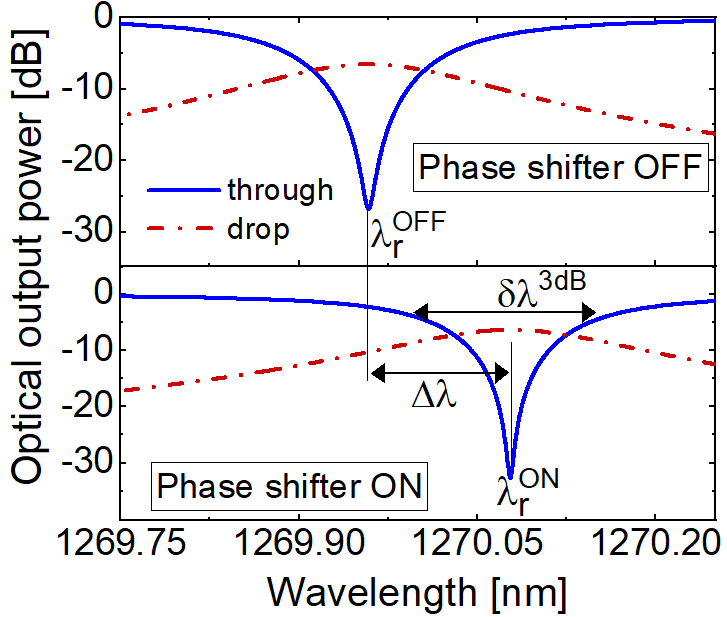}\label{ringtranmission}}
\end{subfigure}
\hspace{0.15cm}
\begin{subfigure}[]
{\includegraphics[scale=0.36]{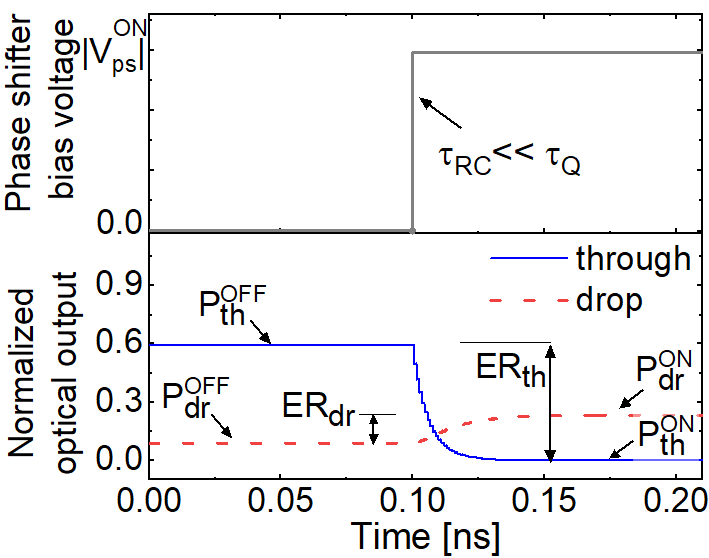}\label{switch}}
\end{subfigure}
%\begin{subfigure}[]
%{\includegraphics[scale=0.35]{Figures/ringdesign.png}\label{ringdesign}}
%\end{subfigure}
\vspace{-0.25cm}
\caption{(a) Schematic representation of a MRR switch operating at $\lambda_r$ with respect to the OFF and ON states of the PD, (b) and (c) are typical wavelength dependent optical transmission and switching  characteristics, respectively at the through- and drop-ports for the OFF state and ON state of the PD.}
\label{ringscheme}
\end{figure*}

From the above discussion, it is clear that the write operation involves more numbers of MRRs and PDs compared to the read operation ($t_\text{write}>t_\text{read}$), and thus the overall speed (or bandwidth) of the memory is specified in terms write speed. The operating speed of our proposed O-SRAM is limited by the bandwidths of MRRs ($BW_{MRR}$) and PDs ($BW_{PD}$) used in the circuit. In general, the MRR bandwidth is expressed using the empirical formula \( 1/BW_{MRR}^2=1/BW_Q^2+1/BW_{RC}^2\) where $BW_{Q}=(2\pi \tau_\text{Q})^{-1}$ is a function of cavity photon life time $\tau_{Q}$ and $BW_{RC}=(2\pi RC)^{-1}$ is a function of RC ($=\tau_{RC}$) time constant of the phase shifter \cite{dong2009low}. At any operating $\lambda$ and a given quality factor $Q$, $\tau_{Q}$ can be expressed as $\tau_{Q}=\lambda Q/(2\pi c)$ where, c is the light speed in vacuum. The $Q$-factor is a critical design parameter of an MRR switch. The optical bandwidth $BW_{Q}$ reduces as $Q$ increases. On the other hand, the photon interaction with the injected/depleted carriers reduces as $Q$ ($\tau_Q$) reduces, and thus large bias voltage is required to switch the MRR between non-resonance to resonance. Thus, a moderate $Q$ factor $\sim 5000-7000$ is preferred for our design, providing a BW $\sim$ 30 GHz to 50 GHz at operating $\lambda\sim1270~$nm. The phase shifter delay, $\tau_{RC}$, is relatively smaller for carrier depletion (p-n) type phase-shifter compared to carrier injection (p-i-n) type phase-shifter \cite{soref1987}. This is because p-i-n devices take additional time delay to sweep out the injected carriers from the intrinsic (i-) region after removing the bias voltage across the junction.

We adapted the ring with SOI waveguide of ~300 nm $\times$ 400 nm, ring radius = 10 $\upmu$m. The waveguide loss, effective index versus bias voltage) and phase-shifter (p-n type) design parameters were adapted as reported in \cite{sun2019128} and designed our add-drop MRR switches near $\lambda=1270~$nm. A schematic of the add-drop MRR switch driven by a PD is shown in Fig. \ref{mrr}. A detailed working principle of MRR switches/modulators can be found elsewhere \cite{bogaerts2012silicon,reed2010silicon,xu2005micrometre,soref1987}. 
We carefully designed the coupling coefficients ($k_{1,2}$) between the ring and bus waveguides for a desired optical switching response at operating wavelength $\lambda$, with the following assumptions: (1) the drop port extinction ratio, $ER_{dr}=|P_{dr}^{ON}-P_{dr}^{OFF}|\geq 3$~dB (where $P_{dr}^{ON,OFF}$ are the drop-port power for the ON and OFF states of MRR) for differential sensing between the \textbf{BIT} and \textbf{BIT*} outputs, (2) through-port extinction ratio, $ER_{th}=|P_{th}^{OFF}-P_{th}^{ON}|\geq20$~dB (where $P_{th}^{ON,OFF}$ are the through-port power for the ON and OFF states of MRR) for better stability (either R1 or R2 is in ON state at any time during the memory operation), and (3) $P_{th}^{ON}$ is negligibly small to keep the corresponding PD in dark current mode (OFF state).
Fig. \ref{ringtranmission} shows the transmission characteristics at the through-port ($P_{th}$) and drop-port ($P_{dr}$) of the MRR switch (R1-R4) for OFF (V$_{ps}=0~V$) and ON ($V_{ps}^{ON}=-4~$V) states of the phase shifter ($V_{ps}$ is the reverse bias voltage across the p-n phase shifter). 
The resonance wavelength ($\lambda_r^\text{OFF}$) is shifted by $\Delta\lambda = 117$ pm to $\lambda_r^\text{ON}\sim1270.06~$nm, with $ER_{th}\approx 23~$dB and $ER_{dr}\approx 4.4~$dB.
The $Q$ factor ($=\lambda_r/\delta \lambda^{3dB}$, where $\delta \lambda ^{3dB}$ is 3-dB band width) of the spectrum is $\approx6500$, which results in $\tau_Q\approx4.38~$ps and $BW_Q\approx36~$GHz. 
We choose the operating wavelength ($\lambda_\text{bit}=1270.08~$nm) slightly above $\lambda_r^\text{ON}$ in order to have smooth transition at the output ports and also to satisfy the above conditions. The switching characteristics at $\lambda_\text{bit}$ is shown in Fig.\ref{switch}. 
The $Q$-limited rise time at the drop port and fall time at the through port are calculated to be $\tau_{r}^{dr}\sim29.5$ ps and $\tau_{f}^{th}\sim14$ ps, respectively.
In this case, the RC limited electrical band width of the phase shifter ($BW_{RC}\sim150~$GHz\cite{sun2019128}) is much higher than that of ring resonator ($BW_Q$). %\st{resulting in $BW_{MRR}\approx BW_{Q}\approx36~$GHz.}
As discussed earlier, R1-R4 are identical MRRs of radii $10~\upmu$m operating at $\lambda_\text{bit}$. Similarly, R5 and R6 are identical MRRs of radii $11.2~\upmu$m, designed to operate at resonant wavelength $\lambda_\text{write}\sim 1269.7~$nm.

Besides the bandwidth, it is also important to consider
reverse saturation current $I_s$ of the p-n phase shifter. For the given parameters in \cite{sun2019128}, $I_s$ is calculated to be $\sim4-10~$nA. Since the MRR and PD are connected in series, one must choose a PD with 
the dark current $I_\text{dark}<<I_s$, in order to avoid the inadvertent switching of the MRR due to dark current while the corresponding PD is OFF. For D1-D4, we choose the Ge based photodiode \cite{de2021high} with a low dark current of $\sim30~$pA ($<<I_s$), responsivity $R=0.043~$A/W and operating speed of $\sim14.5~$GHz at $\lambda\sim1.27~\upmu$m.

\begin{figure}[!h]
\centering
{\includegraphics[scale=0.38]{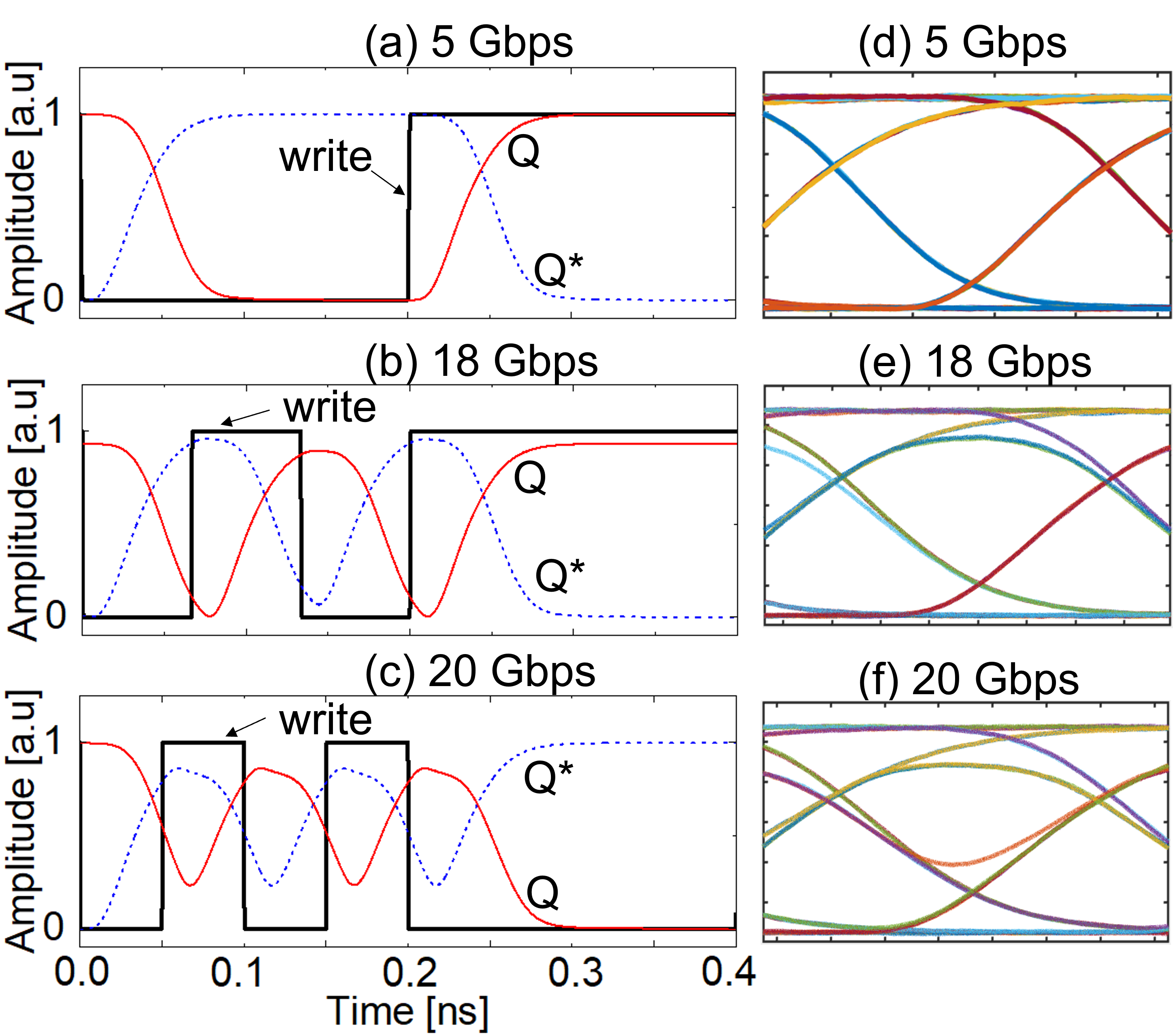}}
\vspace{-0.3cm}
\caption{Simulated switching characteristics (Q and Q*) and corresponding eye-diagrams (calculated at \textbf{Q} output) of the O-SRAM when a `write' pulse at different data rates is applied at the BIT/BIT* waveguide.}
\label{speed}
\vspace{-0.1cm}
\end{figure}
We have simulated and verified the memory operations in \textit{Lumerical Intrerconnect}. The simulated memory write operation at different data rates (5 Gbps, 18 Gbps and 20 Gbps) are shown in Fig. \ref{speed} (a)-(c) and corresponding eye-diagrams (calculated at \textbf{Q} output) in (d)-(f). The input non-return to zero (NRZ) write  signal and the corresponding memory states (Q and Q* outputs) are shown for comparison.
Note that, the outputs reach stable states within the pulse period when operating at low data rates, for example, at 5 Gbps as shown in Fig. \ref{speed} (a). The corresponding rise-time (10\%-90\%) and fall-time (90\%-10\%) are calculated as $t_r\sim46~$ps and $t_f\sim43~$ps, respectively. This means that, the maximum operating speed of the O-SRAM cell is expected to be $1/t_r\approx22~$Gbps. However, given the frequency response of the PD, at higher data rate the outputs of the photodiodes do not reach its maximum value and hence MRRs, that are driven by PDs, operate at slightly lower bias voltage, which in turn changes the operating conditions (i.e, $\Delta\lambda$, $ER_{dr}$, $ER_{th}$, etc.). 
Thus, the actual operating speed of the memory is slightly lower than the calculated value of $1/t_r\approx22~$Gbps. Through simulations, we noticed that the output states (\textbf{Q} and \textbf{Q*}) are distinguishable with clear eye opening when the memory operates $\leq 20~$Gbps (NRZ). This has been shown in Fig. \ref{speed}(b) \& (e)  for 18 Gbps and (c) \& (f) for 20 Gbps. Thus, the maximum speed of the memory is $\sim20~$Gbps. It is worth mentioning, that these initial simulations are based on published literature for photodiodes \cite{de2021high} and MRR modulators \cite{sun2019128} that have been optimized for telecommunication applications. The preliminary speed, required bias voltage and other metrics are, therefore limited by the reported device metrics. Significant improvements can be achieved by device optimization of MRRs and PDs designed specifically for memory storage.

%%%%%%%%%%%%%%%%%%%%%%% 2-COLUMN TABLE %%%%%%%%%%%%%%%%%%%%%%
\begin{table}[!h]
\vspace{-0.2cm}
\caption{Performance comparison of various integrated optical-SRAMs. **The reported switching speed and energy are based on simulation parameters from published literature for photodiodes \cite{de2021high} and MRR modulators
\cite{sun2019128}. *Flip flop only. }
\vspace{-0.5cm}
%\footnotesize
\label{comparison}
\begin{center}
\begin{tabular}{|p{2.3cm}|c|c|c|c|c|p{1.35cm}|}
 \hline
 \multirow{3}{2.3cm}{Device} & \multirow{3}{0.8cm}{Speed (Gbps)} & \multicolumn{4}{c|}{Energy efficiency (pJ/bit)} & \multirow{3}{1.35cm}{Footprint ($\mu \text{m}^2$)}\\
 \cline{3-6}
& &\multicolumn{2}{c|}{Static} &\multicolumn{2}{c|}{Switching} &  \\
 \cline{3-6}
& & Elec. & Opt. &Elec & Opt. &  \\
\hline
 SOA-MZIs \cite{pleros2008optical,liu2006packaged} & 5& 120& $-$ & $-$ & 0.6  & 540$\times10^6$ \\ 
  \hline
  SOA-MZIs \cite{pitris2016wdm}  & 5& 120&  $-$ & $-$ &3  &  12$\times10^6$   \\ 
  \hline
SOA-MZIs  \cite{tsakyridis201910} & 10 & 180& $-$ & $-$ & 0.5  & 12$\times$10$^6$  \\ 
  \hline
 SOA-Ring laser \cite{li2009optical} &1 & $-$ & $-$ &  $-$&  $-$  & 475$\times$400*   \\ 
  \hline
 Photonic crystal \cite{alexoudi2016iii} & 10 & $-$ & 0.01&  $-$& 0.028  & 6.2*   \\ 
  \hline
  Optomechanic \cite{dong2015nano} & 8 &  $-$& $-$ & $-$ &  $-$  & 100$\times$100* \\ 
  \hline
 This work** & 20 &2.5e-6 &1.67e-5 &1.04 &  3.5e-5 & 60$\times$40 \\ 
  \hline
\end{tabular}
\end{center}
\vspace{-0.1cm}
\end{table}
\normalsize
%%%%%%%%%%%%%%%%%%%%%%% 2 COLUMN TABLE %%%%%%%%%%%%%%%%%%%%%%

Table \ref{comparison} shows the comparison of performances of various optical SRAMs reported in the literature. The static energy consumption of our device is estimated as $V_\text{bias}\times(I_{s}+3I_\text{dark})\times t_{bit}\approx 2.5~$aJ/bit, where we assume only one of the PD is ON, and the remaining are OFF at any time during the static operation. Similarly, the static optical energy is calculated as the fraction of internal laser power ($P_\text{laser}$) to the memory speed. With the given PD parameters, we calculated the minimum optical input power to the PD (say D1) as $\approx100~$nW which ensures the MRR (say R1) is in ON state. This corresponds to total internal laser power, $P_\text{laser}=334~$nW, considering the power associated with 3-dB power splitter PS1 (refer Fig. \ref{layout1}) and the transfer function of MRR (R1 or R2) from Fig \ref{ringtranmission}.
%From Fig. \ref{ringtranmission}, the minimum input to the optical input to the MRR is $\sim167$~nW, and hence $P_\text{laser}=334~$nW. 
Thus, the static optical energy consumption is $\sim16.7~$aJ/bit. Since all six MRRs are involved in the write operations, the dynamic electrical energy consumption is $6\times CV_\text{bias}^2/4=1.04~$pJ/bit, where $C=45~$fF \cite{sun2019128}. Similarly, the dynamic optical energy is the sum of all three lasers involved in the operation, i.e., internal laser source, write/read enable laser, and write laser. This is calculated as $\sim40~$aJ/bit. Note, the estimated power budget does not include energy needed for thermal detuning of the MRRs. In general,  the MRR operating wavelength is detuned using thermal phase-shifters to compensate for any wavelength shift due to process variations \cite{sun2019128,padmaraju2014resolving}. Such thermal detuning could be implemented on global (for all MRRs in a memory array) or local (for individual MRRs) level based on the scale and nature of process variations associated with specific manufacturing technology.
%Note that we have not included any thermal bias power in the above calculations. In general, the performance of MRR based devices is sensitive to temperature and fabrication variations.  Thus, in practical circuits, the MRR operating wavelength is detuned using thermal phase-shifters and it compensates for the wavelength shift due to device parameter variations.  From Fig. \ref{ringscheme} (d), we estimated that the device could be operated in a stable state within $\approx\lambda_r^{on}\pm15~$pm. Resonance wavelength shift $>15~$pm implies that the state of memory is unpredictable.  If this shift is due to global process variations (assuming identical change in all MRRs), one can bring the memory to a stable state by detuning the (internal laser) operating wavelength to the resonance wavelength of the MRRs ($\lambda_r^{on}$).  However, for this designed O-SRAM that is dependent on the chosen device components, wavelength shifts greater than 15 pm due to local process variations can only be addressed with integrated microheaters or with highly fabrication tolerant design. 

%\begin{figure}[t]
%\centering
%{\includegraphics[scale=0.4]{Figures/array1.png}}
%\caption{O-SRAM cells in arrange in a 2D array. The `word' line waveguides %are shared in horizontal direction and the BIT, BIT* waveguides are shared %in vertical direction, resulting in a compact array layout. }
%\label{array}
%\end{figure}

In conclusion, we have designed and simulated an O-SRAM cell using cross-coupled MRRs and PDs. The performance of the proposed memory is simulated based on published literature for MRR modulators \cite{sun2019128} and PDs \cite{de2021high}. The operating speed  and static (switching) energy consumption are estimated as $20~$Gb/s and $\sim16.7~$aJ/bit ($\sim1.04$~pJ/bit), respectively. The overall footprint of the device is $\sim2400~\mu$m$^2$. Finally, we would like to highlight some key prospects of our O-SRAM; 1) the proposed optical memory is amenable to large-scale manufacturing in existing silicon photonics foundry process without need of explicit material/process modifications, 2) since optical signals can travel long distances with minimal loss, the O-SRAM bit-cell can be replicated along rows and columns to create very-large scale, ultra-high speed memory arrays, 3) the functional similarity of proposed O-SRAM with electrical SRAM opens up new pathways to implement emerging paradigms of in-memory and near-memory computing, similar to their electrical counterparts, within large-scale optical memory arrays. Thus, the proposed O-SRAM bit-cell can be used to create ultra-fast, ultra-large scale on-chip memory system - a key component required for achieving the holy grail of general purpose optical computing. 

\begin{acknowledgments}
\vspace{-1em}
The authors would like to thank Dr. Michal Rakowski from GlobalFoundries for fruitful technical discussions. This work was supported in part by Mousetrap fund at University of Southern California.
\end{acknowledgments}

\vspace{-1em}
\section*{AUTHOR DECLARATIONS}
\vspace{-1em}
\subsection*{Conflict of Interest}
\vspace{-1em}
The authors report no conflicts of interest.
\vspace{-1em}
\section*{DATA AVAILABILITY}
\vspace{-1em}
The data that supports the findings of this study are available within the article.

%\nocite{*}
\bibliography{Opticalsram}% Produces the bibliography via BibTeX.

%merlin.mbs aipnum4-1.bst 2010-07-25 4.21a (PWD, AO, DPC) hacked
%Control: key (0)
%Control: author (8) initials jnrlst
%Control: editor formatted (1) identically to author
%Control: production of article title (0) allowed
%Control: page (1) range
%Control: year (1) truncated
%Control: production of eprint (0) enabled
\begin{thebibliography}{26}%
\makeatletter
\providecommand \@ifxundefined [1]{%
 \@ifx{#1\undefined}
}%
\providecommand \@ifnum [1]{%
 \ifnum #1\expandafter \@firstoftwo
 \else \expandafter \@secondoftwo
 \fi
}%
\providecommand \@ifx [1]{%
 \ifx #1\expandafter \@firstoftwo
 \else \expandafter \@secondoftwo
 \fi
}%
\providecommand \natexlab [1]{#1}%
\providecommand \enquote  [1]{``#1''}%
\providecommand \bibnamefont  [1]{#1}%
\providecommand \bibfnamefont [1]{#1}%
\providecommand \citenamefont [1]{#1}%
\providecommand \href@noop [0]{\@secondoftwo}%
\providecommand \href [0]{\begingroup \@sanitize@url \@href}%
\providecommand \@href[1]{\@@startlink{#1}\@@href}%
\providecommand \@@href[1]{\endgroup#1\@@endlink}%
\providecommand \@sanitize@url [0]{\catcode `\\12\catcode `\$12\catcode
  `\&12\catcode `\#12\catcode `\^12\catcode `\_12\catcode `\%12\relax}%
\providecommand \@@startlink[1]{}%
\providecommand \@@endlink[0]{}%
\providecommand \url  [0]{\begingroup\@sanitize@url \@url }%
\providecommand \@url [1]{\endgroup\@href {#1}{\urlprefix }}%
\providecommand \urlprefix  [0]{URL }%
\providecommand \Eprint [0]{\href }%
\providecommand \doibase [0]{http://dx.doi.org/}%
\providecommand \selectlanguage [0]{\@gobble}%
\providecommand \bibinfo  [0]{\@secondoftwo}%
\providecommand \bibfield  [0]{\@secondoftwo}%
\providecommand \translation [1]{[#1]}%
\providecommand \BibitemOpen [0]{}%
\providecommand \bibitemStop [0]{}%
\providecommand \bibitemNoStop [0]{.\EOS\space}%
\providecommand \EOS [0]{\spacefactor3000\relax}%
\providecommand \BibitemShut  [1]{\csname bibitem#1\endcsname}%
\let\auto@bib@innerbib\@empty
%</preamble>
\bibitem [{\citenamefont {Khan}, \citenamefont {Hounshell},\ and\ \citenamefont
  {Fuchs}(2018)}]{khan2018science}%
  \BibitemOpen
  \bibfield  {author} {\bibinfo {author} {\bibfnamefont {H.~N.}\ \bibnamefont
  {Khan}}, \bibinfo {author} {\bibfnamefont {D.~A.}\ \bibnamefont {Hounshell}},
  \ and\ \bibinfo {author} {\bibfnamefont {E.~R.}\ \bibnamefont {Fuchs}},\
  }\bibfield  {title} {\enquote {\bibinfo {title} {Science and research policy
  at the end of moore’s law},}\ }\href@noop {} {\bibfield  {journal}
  {\bibinfo  {journal} {Nature Electronics}\ }\textbf {\bibinfo {volume} {1}},\
  \bibinfo {pages} {14--21} (\bibinfo {year} {2018})}\BibitemShut {NoStop}%
\bibitem [{\citenamefont {Ahmed}\ and\ \citenamefont
  {Wahed}(2020)}]{ahmed2020democratization}%
  \BibitemOpen
  \bibfield  {author} {\bibinfo {author} {\bibfnamefont {N.}~\bibnamefont
  {Ahmed}}\ and\ \bibinfo {author} {\bibfnamefont {M.}~\bibnamefont {Wahed}},\
  }\bibfield  {title} {\enquote {\bibinfo {title} {The de-democratization of
  {AI}: Deep learning and the compute divide in artificial intelligence
  research},}\ }\href@noop {} {\bibfield  {journal} {\bibinfo  {journal} {arXiv
  preprint arXiv:2010.15581}\ } (\bibinfo {year} {2020})}\BibitemShut {NoStop}%
\bibitem [{\citenamefont {Donofrio}\ \emph {et~al.}(2009)\citenamefont
  {Donofrio}, \citenamefont {Oliker}, \citenamefont {Shalf}, \citenamefont
  {Wehner}, \citenamefont {Rowen}, \citenamefont {Krueger}, \citenamefont
  {Kamil},\ and\ \citenamefont {Mohiyuddin}}]{donofrio2009energy}%
  \BibitemOpen
  \bibfield  {author} {\bibinfo {author} {\bibfnamefont {D.}~\bibnamefont
  {Donofrio}}, \bibinfo {author} {\bibfnamefont {L.}~\bibnamefont {Oliker}},
  \bibinfo {author} {\bibfnamefont {J.}~\bibnamefont {Shalf}}, \bibinfo
  {author} {\bibfnamefont {M.~F.}\ \bibnamefont {Wehner}}, \bibinfo {author}
  {\bibfnamefont {C.}~\bibnamefont {Rowen}}, \bibinfo {author} {\bibfnamefont
  {J.}~\bibnamefont {Krueger}}, \bibinfo {author} {\bibfnamefont
  {S.}~\bibnamefont {Kamil}}, \ and\ \bibinfo {author} {\bibfnamefont
  {M.}~\bibnamefont {Mohiyuddin}},\ }\bibfield  {title} {\enquote {\bibinfo
  {title} {Energy-efficient computing for extreme-scale science},}\ }\href@noop
  {} {\bibfield  {journal} {\bibinfo  {journal} {Computer}\ }\textbf {\bibinfo
  {volume} {42}},\ \bibinfo {pages} {62--71} (\bibinfo {year}
  {2009})}\BibitemShut {NoStop}%
\bibitem [{\citenamefont {Ambs}(2010)}]{ambs2010optical}%
  \BibitemOpen
  \bibfield  {author} {\bibinfo {author} {\bibfnamefont {P.}~\bibnamefont
  {Ambs}},\ }\bibfield  {title} {\enquote {\bibinfo {title} {Optical computing:
  A 60-year adventure.}}\ }\href@noop {} {\bibfield  {journal} {\bibinfo
  {journal} {Advances in Optical Technologies}\ } (\bibinfo {year}
  {2010})}\BibitemShut {NoStop}%
\bibitem [{\citenamefont {Sawchuk}\ and\ \citenamefont
  {Strand}(1984)}]{sawchuk1984digital}%
  \BibitemOpen
  \bibfield  {author} {\bibinfo {author} {\bibfnamefont {A.~A.}\ \bibnamefont
  {Sawchuk}}\ and\ \bibinfo {author} {\bibfnamefont {T.~C.}\ \bibnamefont
  {Strand}},\ }\bibfield  {title} {\enquote {\bibinfo {title} {Digital optical
  computing},}\ }\href@noop {} {\bibfield  {journal} {\bibinfo  {journal}
  {Proceedings of the IEEE}\ }\textbf {\bibinfo {volume} {72}},\ \bibinfo
  {pages} {758--779} (\bibinfo {year} {1984})}\BibitemShut {NoStop}%
\bibitem [{\citenamefont {McKee}(2004)}]{mckee2004reflections}%
  \BibitemOpen
  \bibfield  {author} {\bibinfo {author} {\bibfnamefont {S.~A.}\ \bibnamefont
  {McKee}},\ }\bibfield  {title} {\enquote {\bibinfo {title} {Reflections on
  the memory wall},}\ }in\ \href@noop {} {\emph {\bibinfo {booktitle}
  {Proceedings of the 1st conference on Computing frontiers}}}\ (\bibinfo
  {year} {2004})\ p.\ \bibinfo {pages} {162}\BibitemShut {NoStop}%
\bibitem [{\citenamefont {Maniotis}\ \emph {et~al.}(2013)\citenamefont
  {Maniotis}, \citenamefont {Fitsios}, \citenamefont {Kanellos},\ and\
  \citenamefont {Pleros}}]{maniotis2013optical}%
  \BibitemOpen
  \bibfield  {author} {\bibinfo {author} {\bibfnamefont {P.}~\bibnamefont
  {Maniotis}}, \bibinfo {author} {\bibfnamefont {D.}~\bibnamefont {Fitsios}},
  \bibinfo {author} {\bibfnamefont {G.}~\bibnamefont {Kanellos}}, \ and\
  \bibinfo {author} {\bibfnamefont {N.}~\bibnamefont {Pleros}},\ }\bibfield
  {title} {\enquote {\bibinfo {title} {Optical buffering for chip
  multiprocessors: a 16ghz optical cache memory architecture},}\ }\href@noop {}
  {\bibfield  {journal} {\bibinfo  {journal} {Journal of Lightwave Technology}\
  }\textbf {\bibinfo {volume} {31}},\ \bibinfo {pages} {4175--4191} (\bibinfo
  {year} {2013})}\BibitemShut {NoStop}%
\bibitem [{\citenamefont {Kodi}\ and\ \citenamefont
  {Louri}(2005)}]{kodi2005design}%
  \BibitemOpen
  \bibfield  {author} {\bibinfo {author} {\bibfnamefont {A.~K.}\ \bibnamefont
  {Kodi}}\ and\ \bibinfo {author} {\bibfnamefont {A.}~\bibnamefont {Louri}},\
  }\bibfield  {title} {\enquote {\bibinfo {title} {Design of a high-speed
  optical interconnect for scalable shared-memory multiprocessors},}\
  }\href@noop {} {\bibfield  {journal} {\bibinfo  {journal} {IEEE micro}\
  }\textbf {\bibinfo {volume} {25}},\ \bibinfo {pages} {41--49} (\bibinfo
  {year} {2005})}\BibitemShut {NoStop}%
\bibitem [{\citenamefont {Brunina}, \citenamefont {Liu},\ and\ \citenamefont
  {Bergman}(2012)}]{brunina2012energy}%
  \BibitemOpen
  \bibfield  {author} {\bibinfo {author} {\bibfnamefont {D.}~\bibnamefont
  {Brunina}}, \bibinfo {author} {\bibfnamefont {D.}~\bibnamefont {Liu}}, \ and\
  \bibinfo {author} {\bibfnamefont {K.}~\bibnamefont {Bergman}},\ }\bibfield
  {title} {\enquote {\bibinfo {title} {An energy-efficient optically connected
  memory module for hybrid packet-and circuit-switched optical networks},}\
  }\href@noop {} {\bibfield  {journal} {\bibinfo  {journal} {IEEE Journal of
  Selected Topics in Quantum Electronics}\ }\textbf {\bibinfo {volume} {19}},\
  \bibinfo {pages} {3700407--3700407} (\bibinfo {year} {2012})}\BibitemShut
  {NoStop}%
\bibitem [{\citenamefont {Yin}\ \emph {et~al.}(2010)\citenamefont {Yin},
  \citenamefont {Proietti}, \citenamefont {Ye}, \citenamefont {Yoo},\ and\
  \citenamefont {Akella}}]{yin2010experimental}%
  \BibitemOpen
  \bibfield  {author} {\bibinfo {author} {\bibfnamefont {Y.}~\bibnamefont
  {Yin}}, \bibinfo {author} {\bibfnamefont {R.}~\bibnamefont {Proietti}},
  \bibinfo {author} {\bibfnamefont {X.}~\bibnamefont {Ye}}, \bibinfo {author}
  {\bibfnamefont {S.~B.}\ \bibnamefont {Yoo}}, \ and\ \bibinfo {author}
  {\bibfnamefont {V.}~\bibnamefont {Akella}},\ }\bibfield  {title} {\enquote
  {\bibinfo {title} {Experimental demonstration of optical processor-memory
  interconnection},}\ }in\ \href@noop {} {\emph {\bibinfo {booktitle} {in Proc.
  AIAI2010, 213-216, Beijing, China, Oct. 2010}}}\ (\bibinfo  {publisher}
  {IET},\ \bibinfo {year} {2010})\BibitemShut {NoStop}%
\bibitem [{\citenamefont {Liu}\ \emph {et~al.}(2006)\citenamefont {Liu},
  \citenamefont {McDougall}, \citenamefont {Hill}, \citenamefont {Maxwell},
  \citenamefont {Zhang}, \citenamefont {Harmon}, \citenamefont {Huijskens},
  \citenamefont {Rivers}, \citenamefont {Dorren},\ and\ \citenamefont
  {Poustie}}]{liu2006packaged}%
  \BibitemOpen
  \bibfield  {author} {\bibinfo {author} {\bibfnamefont {Y.}~\bibnamefont
  {Liu}}, \bibinfo {author} {\bibfnamefont {R.}~\bibnamefont {McDougall}},
  \bibinfo {author} {\bibfnamefont {M.}~\bibnamefont {Hill}}, \bibinfo {author}
  {\bibfnamefont {G.}~\bibnamefont {Maxwell}}, \bibinfo {author} {\bibfnamefont
  {S.}~\bibnamefont {Zhang}}, \bibinfo {author} {\bibfnamefont
  {R.}~\bibnamefont {Harmon}}, \bibinfo {author} {\bibfnamefont
  {F.}~\bibnamefont {Huijskens}}, \bibinfo {author} {\bibfnamefont
  {L.}~\bibnamefont {Rivers}}, \bibinfo {author} {\bibfnamefont
  {H.}~\bibnamefont {Dorren}}, \ and\ \bibinfo {author} {\bibfnamefont
  {A.}~\bibnamefont {Poustie}},\ }\bibfield  {title} {\enquote {\bibinfo
  {title} {Packaged and hybrid integrated all-optical flip-flop memory},}\
  }\href@noop {} {\bibfield  {journal} {\bibinfo  {journal} {Electronics
  Letters}\ }\textbf {\bibinfo {volume} {42}},\ \bibinfo {pages} {1399--1400}
  (\bibinfo {year} {2006})}\BibitemShut {NoStop}%
\bibitem [{\citenamefont {Pleros}\ \emph {et~al.}(2008)\citenamefont {Pleros},
  \citenamefont {Apostolopoulos}, \citenamefont {Petrantonakis}, \citenamefont
  {Stamatiadis},\ and\ \citenamefont {Avramopoulos}}]{pleros2008optical}%
  \BibitemOpen
  \bibfield  {author} {\bibinfo {author} {\bibfnamefont {N.}~\bibnamefont
  {Pleros}}, \bibinfo {author} {\bibfnamefont {D.}~\bibnamefont
  {Apostolopoulos}}, \bibinfo {author} {\bibfnamefont {D.}~\bibnamefont
  {Petrantonakis}}, \bibinfo {author} {\bibfnamefont {C.}~\bibnamefont
  {Stamatiadis}}, \ and\ \bibinfo {author} {\bibfnamefont {H.}~\bibnamefont
  {Avramopoulos}},\ }\bibfield  {title} {\enquote {\bibinfo {title} {Optical
  static {RAM} cell},}\ }\href@noop {} {\bibfield  {journal} {\bibinfo
  {journal} {IEEE Photonics Technology Letters}\ }\textbf {\bibinfo {volume}
  {21}},\ \bibinfo {pages} {73--75} (\bibinfo {year} {2008})}\BibitemShut
  {NoStop}%
\bibitem [{\citenamefont {Pitris}\ \emph {et~al.}(2016)\citenamefont {Pitris},
  \citenamefont {Vagionas}, \citenamefont {Tekin}, \citenamefont {Broeke},
  \citenamefont {Kanellos},\ and\ \citenamefont {Pleros}}]{pitris2016wdm}%
  \BibitemOpen
  \bibfield  {author} {\bibinfo {author} {\bibfnamefont {S.}~\bibnamefont
  {Pitris}}, \bibinfo {author} {\bibfnamefont {C.}~\bibnamefont {Vagionas}},
  \bibinfo {author} {\bibfnamefont {T.}~\bibnamefont {Tekin}}, \bibinfo
  {author} {\bibfnamefont {R.}~\bibnamefont {Broeke}}, \bibinfo {author}
  {\bibfnamefont {G.}~\bibnamefont {Kanellos}}, \ and\ \bibinfo {author}
  {\bibfnamefont {N.}~\bibnamefont {Pleros}},\ }\bibfield  {title} {\enquote
  {\bibinfo {title} {{WDM}-enabled optical {RAM} at 5 {G}b/s using a monolithic
  inp flip-flop chip},}\ }\href@noop {} {\bibfield  {journal} {\bibinfo
  {journal} {IEEE Photonics {J}ournal}\ }\textbf {\bibinfo {volume} {8}},\
  \bibinfo {pages} {1--7} (\bibinfo {year} {2016})}\BibitemShut {NoStop}%
\bibitem [{\citenamefont {Tsakyridis}\ \emph {et~al.}(2019)\citenamefont
  {Tsakyridis}, \citenamefont {Alexoudi}, \citenamefont {Miliou}, \citenamefont
  {Pleros},\ and\ \citenamefont {Vagionas}}]{tsakyridis201910}%
  \BibitemOpen
  \bibfield  {author} {\bibinfo {author} {\bibfnamefont {A.}~\bibnamefont
  {Tsakyridis}}, \bibinfo {author} {\bibfnamefont {T.}~\bibnamefont
  {Alexoudi}}, \bibinfo {author} {\bibfnamefont {A.}~\bibnamefont {Miliou}},
  \bibinfo {author} {\bibfnamefont {N.}~\bibnamefont {Pleros}}, \ and\ \bibinfo
  {author} {\bibfnamefont {C.}~\bibnamefont {Vagionas}},\ }\bibfield  {title}
  {\enquote {\bibinfo {title} {10 {G}b/s optical random access memory ({RAM})
  cell},}\ }\href@noop {} {\bibfield  {journal} {\bibinfo  {journal} {Optics
  {L}etters}\ }\textbf {\bibinfo {volume} {44}},\ \bibinfo {pages} {1821--1824}
  (\bibinfo {year} {2019})}\BibitemShut {NoStop}%
\bibitem [{\citenamefont {Li}\ \emph {et~al.}(2009)\citenamefont {Li},
  \citenamefont {Memon}, \citenamefont {Mezosi}, \citenamefont {Wang},
  \citenamefont {Sorel},\ and\ \citenamefont {Yu}}]{li2009optical}%
  \BibitemOpen
  \bibfield  {author} {\bibinfo {author} {\bibfnamefont {B.}~\bibnamefont
  {Li}}, \bibinfo {author} {\bibfnamefont {M.~I.}\ \bibnamefont {Memon}},
  \bibinfo {author} {\bibfnamefont {G.}~\bibnamefont {Mezosi}}, \bibinfo
  {author} {\bibfnamefont {Z.}~\bibnamefont {Wang}}, \bibinfo {author}
  {\bibfnamefont {M.}~\bibnamefont {Sorel}}, \ and\ \bibinfo {author}
  {\bibfnamefont {S.}~\bibnamefont {Yu}},\ }\bibfield  {title} {\enquote
  {\bibinfo {title} {Optical static random access memory cell using an
  integrated semiconductor ring laser},}\ }in\ \href@noop {} {\emph {\bibinfo
  {booktitle} {2009 International Conference on Photonics in Switching}}}\
  (\bibinfo {organization} {IEEE},\ \bibinfo {year} {2009})\ pp.\ \bibinfo
  {pages} {1--2}\BibitemShut {NoStop}%
\bibitem [{\citenamefont {Alexoudi}\ \emph {et~al.}(2016)\citenamefont
  {Alexoudi}, \citenamefont {Fitsios}, \citenamefont {Bazin}, \citenamefont
  {Monnier}, \citenamefont {Raj}, \citenamefont {Miliou}, \citenamefont
  {Kanellos}, \citenamefont {Pleros},\ and\ \citenamefont
  {Raineri}}]{alexoudi2016iii}%
  \BibitemOpen
  \bibfield  {author} {\bibinfo {author} {\bibfnamefont {T.}~\bibnamefont
  {Alexoudi}}, \bibinfo {author} {\bibfnamefont {D.}~\bibnamefont {Fitsios}},
  \bibinfo {author} {\bibfnamefont {A.}~\bibnamefont {Bazin}}, \bibinfo
  {author} {\bibfnamefont {P.}~\bibnamefont {Monnier}}, \bibinfo {author}
  {\bibfnamefont {R.}~\bibnamefont {Raj}}, \bibinfo {author} {\bibfnamefont
  {A.}~\bibnamefont {Miliou}}, \bibinfo {author} {\bibfnamefont {G.~T.}\
  \bibnamefont {Kanellos}}, \bibinfo {author} {\bibfnamefont {N.}~\bibnamefont
  {Pleros}}, \ and\ \bibinfo {author} {\bibfnamefont {F.}~\bibnamefont
  {Raineri}},\ }\bibfield  {title} {\enquote {\bibinfo {title}
  {{III}--{V}-on-{S}i photonic crystal nanocavity laser technology for optical
  static random access memories},}\ }\href@noop {} {\bibfield  {journal}
  {\bibinfo  {journal} {IEEE Journal of Selected Topics in Quantum
  Electronics}\ }\textbf {\bibinfo {volume} {22}},\ \bibinfo {pages} {295--304}
  (\bibinfo {year} {2016})}\BibitemShut {NoStop}%
\bibitem [{\citenamefont {Dong}\ \emph {et~al.}(2015)\citenamefont {Dong},
  \citenamefont {Cai}, \citenamefont {Gu}, \citenamefont {Yang}, \citenamefont
  {Jin}, \citenamefont {Hao}, \citenamefont {Kwong},\ and\ \citenamefont
  {Liu}}]{dong2015nano}%
  \BibitemOpen
  \bibfield  {author} {\bibinfo {author} {\bibfnamefont {B.}~\bibnamefont
  {Dong}}, \bibinfo {author} {\bibfnamefont {H.}~\bibnamefont {Cai}}, \bibinfo
  {author} {\bibfnamefont {Y.}~\bibnamefont {Gu}}, \bibinfo {author}
  {\bibfnamefont {Z.}~\bibnamefont {Yang}}, \bibinfo {author} {\bibfnamefont
  {Y.}~\bibnamefont {Jin}}, \bibinfo {author} {\bibfnamefont {Y.}~\bibnamefont
  {Hao}}, \bibinfo {author} {\bibfnamefont {D.}~\bibnamefont {Kwong}}, \ and\
  \bibinfo {author} {\bibfnamefont {A.}~\bibnamefont {Liu}},\ }\bibfield
  {title} {\enquote {\bibinfo {title} {Nano-optomechanical static random access
  memory ({SRAM})},}\ }in\ \href@noop {} {\emph {\bibinfo {booktitle} {2015
  28th IEEE International Conference on Micro Electro Mechanical Systems
  (MEMS)}}}\ (\bibinfo {organization} {IEEE},\ \bibinfo {year} {2015})\ pp.\
  \bibinfo {pages} {49--52}\BibitemShut {NoStop}%
\bibitem [{\citenamefont {Mathur}\ \emph {et~al.}(2020)\citenamefont {Mathur},
  \citenamefont {Bhargava}, \citenamefont {Salahuddin}, \citenamefont
  {Schuddinck}, \citenamefont {Ryckaert}, \citenamefont {Annamalai},
  \citenamefont {Gupta}, \citenamefont {Chong}, \citenamefont {Sinha},
  \citenamefont {Cline} \emph {et~al.}}]{mathur2020buried}%
  \BibitemOpen
  \bibfield  {author} {\bibinfo {author} {\bibfnamefont {R.}~\bibnamefont
  {Mathur}}, \bibinfo {author} {\bibfnamefont {M.}~\bibnamefont {Bhargava}},
  \bibinfo {author} {\bibfnamefont {S.}~\bibnamefont {Salahuddin}}, \bibinfo
  {author} {\bibfnamefont {P.}~\bibnamefont {Schuddinck}}, \bibinfo {author}
  {\bibfnamefont {J.}~\bibnamefont {Ryckaert}}, \bibinfo {author}
  {\bibfnamefont {S.}~\bibnamefont {Annamalai}}, \bibinfo {author}
  {\bibfnamefont {A.}~\bibnamefont {Gupta}}, \bibinfo {author} {\bibfnamefont
  {Y.}~\bibnamefont {Chong}}, \bibinfo {author} {\bibfnamefont
  {S.}~\bibnamefont {Sinha}}, \bibinfo {author} {\bibfnamefont
  {B.}~\bibnamefont {Cline}},  \emph {et~al.},\ }\bibfield  {title} {\enquote
  {\bibinfo {title} {Buried bitline for sub-5nm sram design},}\ }in\ \href@noop
  {} {\emph {\bibinfo {booktitle} {2020 IEEE International Electron Devices
  Meeting (IEDM)}}}\ (\bibinfo {organization} {IEEE},\ \bibinfo {year} {2020})\
  pp.\ \bibinfo {pages} {20--2}\BibitemShut {NoStop}%
\bibitem [{\citenamefont {Dong}\ \emph {et~al.}(2009)\citenamefont {Dong},
  \citenamefont {Liao}, \citenamefont {Feng}, \citenamefont {Liang},
  \citenamefont {Zheng}, \citenamefont {Shafiiha}, \citenamefont {Kung},
  \citenamefont {Qian}, \citenamefont {Li}, \citenamefont {Zheng} \emph
  {et~al.}}]{dong2009low}%
  \BibitemOpen
  \bibfield  {author} {\bibinfo {author} {\bibfnamefont {P.}~\bibnamefont
  {Dong}}, \bibinfo {author} {\bibfnamefont {S.}~\bibnamefont {Liao}}, \bibinfo
  {author} {\bibfnamefont {D.}~\bibnamefont {Feng}}, \bibinfo {author}
  {\bibfnamefont {H.}~\bibnamefont {Liang}}, \bibinfo {author} {\bibfnamefont
  {D.}~\bibnamefont {Zheng}}, \bibinfo {author} {\bibfnamefont
  {R.}~\bibnamefont {Shafiiha}}, \bibinfo {author} {\bibfnamefont {C.-C.}\
  \bibnamefont {Kung}}, \bibinfo {author} {\bibfnamefont {W.}~\bibnamefont
  {Qian}}, \bibinfo {author} {\bibfnamefont {G.}~\bibnamefont {Li}}, \bibinfo
  {author} {\bibfnamefont {X.}~\bibnamefont {Zheng}},  \emph {et~al.},\
  }\bibfield  {title} {\enquote {\bibinfo {title} {Low {V}$_\text{pp}$,
  ultralow-energy, compact, high-speed silicon electro-optic modulator},}\
  }\href@noop {} {\bibfield  {journal} {\bibinfo  {journal} {Optics {E}xpress}\
  }\textbf {\bibinfo {volume} {17}},\ \bibinfo {pages} {22484--22490} (\bibinfo
  {year} {2009})}\BibitemShut {NoStop}%
\bibitem [{\citenamefont {Soref}\ and\ \citenamefont
  {Bennett}(1987)}]{soref1987}%
  \BibitemOpen
  \bibfield  {author} {\bibinfo {author} {\bibfnamefont {R.}~\bibnamefont
  {Soref}}\ and\ \bibinfo {author} {\bibfnamefont {B.}~\bibnamefont
  {Bennett}},\ }\bibfield  {title} {\enquote {\bibinfo {title} {Electrooptical
  effects in silicon},}\ }\href@noop {} {\bibfield  {journal} {\bibinfo
  {journal} {IEEE {J}ournal of {Q}uantum {E}lectronics}\ }\textbf {\bibinfo
  {volume} {23}},\ \bibinfo {pages} {123--129} (\bibinfo {year}
  {1987})}\BibitemShut {NoStop}%
\bibitem [{\citenamefont {Sun}\ \emph {et~al.}(2019)\citenamefont {Sun},
  \citenamefont {Kumar}, \citenamefont {Sakib}, \citenamefont {Driscoll},
  \citenamefont {Jayatilleka},\ and\ \citenamefont {Rong}}]{sun2019128}%
  \BibitemOpen
  \bibfield  {author} {\bibinfo {author} {\bibfnamefont {J.}~\bibnamefont
  {Sun}}, \bibinfo {author} {\bibfnamefont {R.}~\bibnamefont {Kumar}}, \bibinfo
  {author} {\bibfnamefont {M.}~\bibnamefont {Sakib}}, \bibinfo {author}
  {\bibfnamefont {J.~B.}\ \bibnamefont {Driscoll}}, \bibinfo {author}
  {\bibfnamefont {H.}~\bibnamefont {Jayatilleka}}, \ and\ \bibinfo {author}
  {\bibfnamefont {H.}~\bibnamefont {Rong}},\ }\bibfield  {title} {\enquote
  {\bibinfo {title} {A 128 {G}b/s pam4 silicon microring modulator with
  integrated thermo-optic resonance tuning},}\ }\href@noop {} {\bibfield
  {journal} {\bibinfo  {journal} {Journal of {L}ightwave {T}echnology}\
  }\textbf {\bibinfo {volume} {37}},\ \bibinfo {pages} {110--115} (\bibinfo
  {year} {2019})}\BibitemShut {NoStop}%
\bibitem [{\citenamefont {Bogaerts}\ \emph {et~al.}(2012)\citenamefont
  {Bogaerts}, \citenamefont {De~Heyn}, \citenamefont {Van~Vaerenbergh},
  \citenamefont {De~Vos}, \citenamefont {Kumar~Selvaraja}, \citenamefont
  {Claes}, \citenamefont {Dumon}, \citenamefont {Bienstman}, \citenamefont
  {Van~Thourhout},\ and\ \citenamefont {Baets}}]{bogaerts2012silicon}%
  \BibitemOpen
  \bibfield  {author} {\bibinfo {author} {\bibfnamefont {W.}~\bibnamefont
  {Bogaerts}}, \bibinfo {author} {\bibfnamefont {P.}~\bibnamefont {De~Heyn}},
  \bibinfo {author} {\bibfnamefont {T.}~\bibnamefont {Van~Vaerenbergh}},
  \bibinfo {author} {\bibfnamefont {K.}~\bibnamefont {De~Vos}}, \bibinfo
  {author} {\bibfnamefont {S.}~\bibnamefont {Kumar~Selvaraja}}, \bibinfo
  {author} {\bibfnamefont {T.}~\bibnamefont {Claes}}, \bibinfo {author}
  {\bibfnamefont {P.}~\bibnamefont {Dumon}}, \bibinfo {author} {\bibfnamefont
  {P.}~\bibnamefont {Bienstman}}, \bibinfo {author} {\bibfnamefont
  {D.}~\bibnamefont {Van~Thourhout}}, \ and\ \bibinfo {author} {\bibfnamefont
  {R.}~\bibnamefont {Baets}},\ }\bibfield  {title} {\enquote {\bibinfo {title}
  {Silicon microring resonators},}\ }\href@noop {} {\bibfield  {journal}
  {\bibinfo  {journal} {Laser \& Photonics Reviews}\ }\textbf {\bibinfo
  {volume} {6}},\ \bibinfo {pages} {47--73} (\bibinfo {year}
  {2012})}\BibitemShut {NoStop}%
\bibitem [{\citenamefont {Reed}\ \emph {et~al.}(2010)\citenamefont {Reed},
  \citenamefont {Mashanovich}, \citenamefont {Gardes},\ and\ \citenamefont
  {Thomson}}]{reed2010silicon}%
  \BibitemOpen
  \bibfield  {author} {\bibinfo {author} {\bibfnamefont {G.~T.}\ \bibnamefont
  {Reed}}, \bibinfo {author} {\bibfnamefont {G.}~\bibnamefont {Mashanovich}},
  \bibinfo {author} {\bibfnamefont {F.~Y.}\ \bibnamefont {Gardes}}, \ and\
  \bibinfo {author} {\bibfnamefont {D.}~\bibnamefont {Thomson}},\ }\bibfield
  {title} {\enquote {\bibinfo {title} {Silicon optical modulators},}\
  }\href@noop {} {\bibfield  {journal} {\bibinfo  {journal} {Nature
  {P}hotonics}\ }\textbf {\bibinfo {volume} {4}},\ \bibinfo {pages} {518--526}
  (\bibinfo {year} {2010})}\BibitemShut {NoStop}%
\bibitem [{\citenamefont {Xu}\ \emph {et~al.}(2005)\citenamefont {Xu},
  \citenamefont {Schmidt}, \citenamefont {Pradhan},\ and\ \citenamefont
  {Lipson}}]{xu2005micrometre}%
  \BibitemOpen
  \bibfield  {author} {\bibinfo {author} {\bibfnamefont {Q.}~\bibnamefont
  {Xu}}, \bibinfo {author} {\bibfnamefont {B.}~\bibnamefont {Schmidt}},
  \bibinfo {author} {\bibfnamefont {S.}~\bibnamefont {Pradhan}}, \ and\
  \bibinfo {author} {\bibfnamefont {M.}~\bibnamefont {Lipson}},\ }\bibfield
  {title} {\enquote {\bibinfo {title} {Micrometre-scale silicon electro-optic
  modulator},}\ }\href@noop {} {\bibfield  {journal} {\bibinfo  {journal}
  {{N}ature}\ }\textbf {\bibinfo {volume} {435}},\ \bibinfo {pages} {325--327}
  (\bibinfo {year} {2005})}\BibitemShut {NoStop}%
\bibitem [{\citenamefont {de~Cea}\ \emph {et~al.}(2021)\citenamefont {de~Cea},
  \citenamefont {Van~Orden}, \citenamefont {Fini}, \citenamefont {Wade},\ and\
  \citenamefont {Ram}}]{de2021high}%
  \BibitemOpen
  \bibfield  {author} {\bibinfo {author} {\bibfnamefont {M.}~\bibnamefont
  {de~Cea}}, \bibinfo {author} {\bibfnamefont {D.}~\bibnamefont {Van~Orden}},
  \bibinfo {author} {\bibfnamefont {J.}~\bibnamefont {Fini}}, \bibinfo {author}
  {\bibfnamefont {M.}~\bibnamefont {Wade}}, \ and\ \bibinfo {author}
  {\bibfnamefont {R.~J.}\ \bibnamefont {Ram}},\ }\bibfield  {title} {\enquote
  {\bibinfo {title} {High-speed, zero-biased silicon-germanium
  photodetector},}\ }\href@noop {} {\bibfield  {journal} {\bibinfo  {journal}
  {APL Photonics}\ }\textbf {\bibinfo {volume} {6}},\ \bibinfo {pages} {041302}
  (\bibinfo {year} {2021})}\BibitemShut {NoStop}%
\bibitem [{\citenamefont {Padmaraju}\ and\ \citenamefont
  {Bergman}(2014)}]{padmaraju2014resolving}%
  \BibitemOpen
  \bibfield  {author} {\bibinfo {author} {\bibfnamefont {K.}~\bibnamefont
  {Padmaraju}}\ and\ \bibinfo {author} {\bibfnamefont {K.}~\bibnamefont
  {Bergman}},\ }\bibfield  {title} {\enquote {\bibinfo {title} {Resolving the
  thermal challenges for silicon microring resonator devices},}\ }\href@noop {}
  {\bibfield  {journal} {\bibinfo  {journal} {Nanophotonics}\ }\textbf
  {\bibinfo {volume} {3}},\ \bibinfo {pages} {269--281} (\bibinfo {year}
  {2014})}\BibitemShut {NoStop}%
\end{thebibliography}%


%merlin.mbs aipnum4-1.bst 2010-07-25 4.21a (PWD, AO, DPC) hacked
%Control: key (0)
%Control: author (8) initials jnrlst
%Control: editor formatted (1) identically to author
%Control: production of article title (0) allowed
%Control: page (1) range
%Control: year (1) truncated
%Control: production of eprint (0) enabled
%

\end{document}